\documentclass[aps,prb,twocolumn,superscriptaddress,showpacs]{revtex4-1}

\usepackage{graphicx}

\newcommand{\tmo}{TbMn$_2$O$_5$}
\newcommand{\rmo}{\textit{R}Mn$_2$O$_5$}
\newcommand{\pmsat}{($0+\delta,~0,~5-\tau$)}
\newcommand{\ppsat}{($0+\delta,~0,~5+\tau$)}

\begin{document}

\title{Magnetically induced electric polarization reversal in multiferroic TbMn$_2$O$_5$: Terbium spin reorientation studied by resonant x-ray diffraction}

\author{R. D. Johnson}\email{r.johnson1@physics.ox.ac.uk}
\affiliation{Department of Physics, Durham University, Rochester Building, South Road, Durham, DH1 3LE, United Kingdom}
\author{C. Mazzoli}
\affiliation{European Synchrotron Radiation Facility, F-38043 Grenoble, France}
\author{S. R. Bland}
\affiliation{Department of Physics, Durham University, Rochester Building, South Road, Durham, DH1 3LE, United Kingdom}
\author{C-H. Du}
\affiliation{Department of Physics, Tamkang University, Tamsui, Taiwan}
\author{P. D. Hatton}
\affiliation{Department of Physics, Durham University, Rochester Building, South Road, Durham, DH1 3LE, United Kingdom}

\date{\today}

\begin{abstract}

In multiferroic \tmo, the behavior of the terbium ions forms a crucial part of the magneto-electric coupling. The result is a magnetically induced reversal of the electric polarization at 2~T. In this article we present the first direct measurement of the terbium magnetic structure under applied magnetic fields. Contrary to the current interpretation of the magnetic properties of \tmo, we show that upon the electric polarization reversal the terbium sub-lattice adopts a canted antiferromagnetic structure with a large component of magnetic moment parallel to the \textit{a}-axis. Furthermore, we provide evidence for a coupling between the manganese $3d$ magnetic structure and the terbium $4f$ magnetism, which is of great significance in the elusive magneto-electric mechanisms at play.

\end{abstract}


\maketitle

\section{Introduction}

Multiferroic materials are likely to be key components in novel data storage and sensing device solutions. Accordingly research in this field has boomed in the last ten years\cite{kimura03,ederer04,spaldin05,cheong07,fiebig05,khomskii09}. The \rmo\ series, in which a variety of magneto-electric effects have been observed, exhibit some of the most extreme multiferroic phenomena in a single phase. A change in the ferroelectricity is observed in response to an applied magnetic field, probably mediated through a change in the magnetic structure or domain occupation. These materials have complex phase diagrams that include a number of commensurate and incommensurate antiferromagnetic phases below $T_\mathrm{N}\sim45$~K (refs. \citenum{tachibana05,radaelli08b,fukunaga10}). In HoMn$_2$O$_5$ a magnetic field applied parallel to the \textit{b}-axis reinstated the electric polarization\cite{kimura06,noda08}. The opposite effect was observed in ErMn$_2$O$_5$, whereby an applied magnetic field suppressed the electric polarization\cite{kimura07_2,noda08}. In TmMn$_2$O$_5$ a magnetic field applied parallel to the \textit{c}-axis induced a polarization flop from the \textit{a}-axis to the \textit{b}-axis in the low temperature incommensurate (LT-ICM) phase\cite{fukunaga09}. Also, in \tmo, a 2~T magnetic field applied parallel to the \textit{a}-axis, also in the LT-ICM phase, resulted in a highly reproducible electric polarization flip, parallel to the \textit{b}-axis\cite{hur04}.

Despite the polarization originating in the manganese magnetic structure\cite{radaelli08b}, the magneto-electric effects in the LT-ICM phase are likely to be due to the behavior of the rare-earth ion sub-lattice. Its large response to the applied magnetic field may cause changes in the symmetry of the system and modifications in the manganese magnetic structure. As such, the rare-earth ions are expected to provide a \textit{magnetic handle} on the system. With such a diverse range of phenomena across the series, the small differences in an otherwise common manganese magnetic structure (shown in the inset of Fig. \ref{histfig}) are unlikely to result in the varied responses to applied magnetic fields. Indeed, in YMn$_2$O$_5$ the manganese magnetic structure was found to be invariant in magnetic fields up to 8~T (ref. \citenum{radaelli08b}), whilst magnetometry measurements of a number of \rmo\ compounds showed the magnetic susceptibility to be dominated by the rare-earth magnetism\cite{alonso97,uga98,hur04}. Further evidence of the importance of the rare-earth ion behavior is that the extreme magneto-electric effects are only evident at low temperatures, often below the spontaneous ordering temperature of the rare-earth sub-lattice.

From a technological standpoint, the repeatable polarization switching evident in \tmo\ is one of the most useful magneto-electric effects that the \rmo\ series displays. An understanding of the behavior of the terbium sub-lattice under applied magnetic fields in the LT-ICM phase is therefore essential. Magnetometry indicated that at 2~T and 2~K, the terbium sub-lattice saturates with an average moment of 8.2~$\mu_\mathrm{B}$/Tb (ref. \citenum{hur04}), close to the theoretical paramagnetic moment of 9.73~$\mu_\mathrm{B}$/Tb. Furthermore, neutron powder diffraction data suggested that the terbium sub-lattice is ferromagnetic above 2.5~T (ref. \citenum{chapon04}). However, a detailed study of the behavior of the terbium magnetic structure under applied magnetic fields has yet to be reported. In this article we present the first ion-specific study of the terbium magnetism in \tmo\ in applied magnetic fields.

\section{Experiment}

\begin{figure}
\includegraphics[width=8cm]{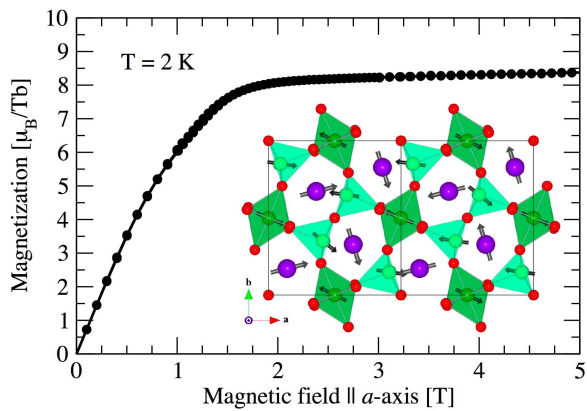}
\caption{\label{histfig}(Color online) The isothermal magnetization of \tmo, measured at 2 K in magnetic fields up to 5 T applied parallel to the \textit{a}-axis. No hysteresis was observed. \textit{Inset}, the commensurate magnetic structure of \tmo\cite{blake05,johnson08}, where Mn$^{4+}$ and Mn$^{3+}$ ions are located within the dark green octahedra and light green square-based pyramid oxygen coordinations, respectively. The terbium ions are shown as purple spheres.}
\end{figure}

A high-quality single crystal of \tmo\ with dimensions approximately 1 x 1 x 0.2~mm$^3$ was grown at the Department of Chemistry in the National Taiwan University by flux growth. The crystal had the (0,~0,~1) Bragg reflection surface-normal to a large facet, which was polished to a mirror-like finish. In order to characterize the sample and check the response to applied magnetic fields at low temperatures, the isothermal magnetization was measured using a Quantum Design MPMS. Fig. \ref{histfig} shows the data collected, measured with the magnetic field applied parallel to the \textit{a}-axis. In agreement with the literature\cite{hur04} the magnetization saturates at approximately 2~T, dominated by the rare-earth moments of 8.2~$\mu_\mathrm{B}$/Tb.

We have performed a resonant x-ray diffraction\cite{hannon88} (RXD) experiment with direct sensitivity to the terbium $5d$ and $4f$ electronic states. RXD has proved to be a lucid probe of the complex magentic structure of \tmo\cite{koo07,okamoto07,johnson08,beale10}. The diffraction measurement was undertaken with an \textit{in-situ} magnetic field of up to 10~T, at beam line ID20, ESRF\cite{ID20}. This enabled us to detect changes in the terbium magnetic structure at the electric polarization reversal. The sample, mounted such that the \textit{c}-axis and \textit{a}-axis were parallel and perpendicular to the scattering plane, respectively, was cooled to 4~K; within the LT-ICM phase. Constraints in the experimental geometry restricted the study to a limited number of magnetic diffraction reflections. Magnetic satellites of the (0,~0,~4), (0,~0,~5) and (0,~0,~6) Bragg reflections were surveyed in the two scattered beam polarization channels parallel ($\pi$') and perpendicular ($\sigma$') to the scattering plane, separated through use of a Au(222) polarization analyzer crystal. For practical reasons the experiment was constrained to the study of the (0+$\delta$,~0,~5-$\tau$) and (0+$\delta$,~0,~5+$\tau$) magnetic satellites, with $\delta\simeq0.487$ and $\tau\simeq0.316$.

\section{Results and discussion}

\begin{figure}
\includegraphics[width=8cm]{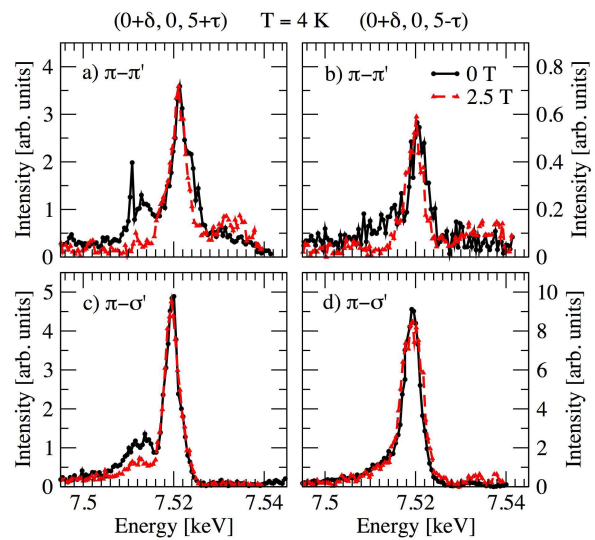}
\caption{\label{escansfig}(Color online) Scans of the scattered intensity as a function of incident x-ray energy through the terbium $L_\mathrm{III}$ edge of (a) \& (c), the \ppsat\ and (b) \& (d), the \pmsat\ reflections, measured in the $\pi'$ and $\sigma'$ polarization channels, respectively. The scans were performed at 4 K, in zero applied magnetic field (black circles) and 2.5 T (red triangles). Note that the intensity units, albeit arbitrary, are scaled equivalently with respect to each other such that a relative comparison between the zero applied field spectra can be made. The 2.5 T spectra are scaled such that the dipole resonance coincides with that of the zero applied field spectra allowing for a clear comparison of the quadrupole resonance behavior.}
\end{figure}

Scans of scattered intensity as a function of incident x-ray energy were performed through the terbium $L_\mathrm{III}$-edge. The scans were then repeated in an applied field of 2.5~T; above the electric polarization reversal transition. The data were corrected for fluctuations in the incident beam intensity and are presented in Fig. \ref{escansfig}. In zero magnetic field, the dipole resonance at 7.520~keV, which probes the terbium $5d$ states, was clearly observed in all scans. As the $5d$ states are de-localized, this resonance is primarily sensitive to the approximately invariant, manganese magnetic structure and hence remains prominent while an external magnetic field is applied. The quadrupole resonance that directly probes the terbium $4f$ states\cite{johnson08} was evident at 7.512~keV in both polarization channels at the (0+$\delta$,~0,~5+$\tau$) reflection. In Fig. \ref{escansfig}, the 2.5~T energy spectra have been scaled such that the dipole resonance coincides with that of the zero field spectra. This allows for a clear comparison of the relative behavior of the quadrupole resonance.

The key result of this study is that at 2~T there occurs a dramatic extinction and suppression of the quadrupole resonance of the (0+$\delta$,~0,~5+$\tau$) reflection in the $\pi'$ and $\sigma'$ polarization channels, respectively. The changes in the relative intensity of the quadrupole resonance induced by the applied magnetic field may occur for two reasons. Either the terbium magnetic structure could dramatically rearrange in response to the applied magnetic field, hence altering the magnetic structure factor, or the terbium incommensurate propagation vector could change with respect to that of the manganese. Later in this article we show that both the dipole and quadrupole resonances have a common magnetic propagation vector. The changes observed in the quadrupole resonance must therefore occur due to a rearrangement of the terbium sub-lattice magnetic structure.

Neutron diffraction data suggested that the terbium ions order ferromagnetically at 2.5~T\cite{chapon04}. This would explain the extinction of the quadrupole resonance measured in the $\pi'$ channel, as in a ferromagnetic structure the magnetic unit cell coincides with the crystallographic unit cell giving zero intensity at all satellite positions. However, the resonance was only partially suppressed in the $\sigma'$ energy spectra. This indicates that the terbium magnetic structure maintains some degree of antiferromagnetic propagation, in common with the manganese magnetism. The terbium sub-lattice therefore adopts a canted antiferromagnetic structure with a large ferromagnetic component in the \textit{a}-axis direction. This explains the reduced terbium magnetic moment evident in the magnetometry. It is not possible to provide a more detailed inference of the magnetic structure in applied field. The zero field manganese magnetic structure would be required as a basis, which is yet to be determined. Structure factor calculations could provide clues about the actual magnetic structure, however many more reflections would need to be measured than were accessible in this experiment.

\begin{figure}
\includegraphics[width=8cm]{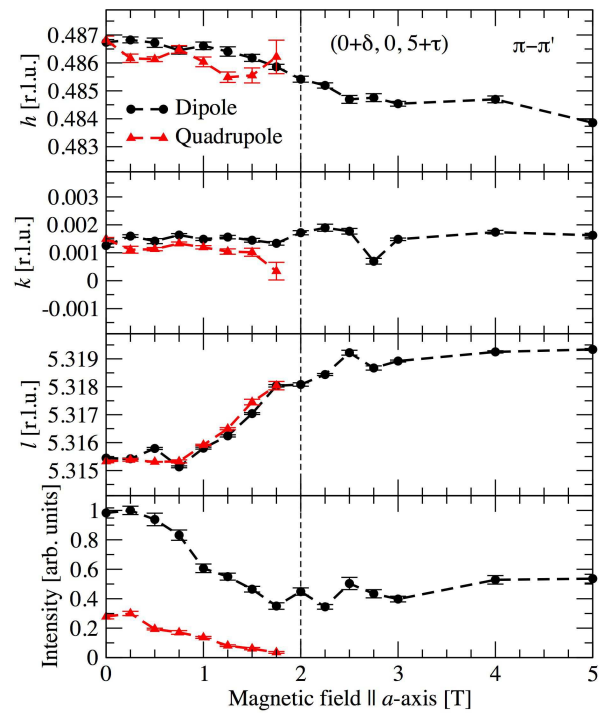}
\caption{\label{fielddepfig}(Color online) The magnetic field dependence of the \ppsat\ wave vector and integrated intensity, measured in the $\pi'$ polarization channel. The dipole and quadrupole resonance dependencies are shown in black circles and red triangles, respectively. The 2 T polarization flip transition is indicated by the vertical dashed line.}
\end{figure}

Fig. \ref{fielddepfig} shows the magnetic field dependence of both dipole and quadrupole, (0+$\delta$,~0,~5+$\tau$) energy resonances, measured in the $\pi'$ polarization channel. The magnetic propagation vector was approximately invariant in the \textit{h} and \textit{k} directions, with only a small monotonic decrease in the \textit{h} direction. This was also found to be the case in the measurement of other magnetic satellite positions, not shown here. By comparison a substantial increase in the \textit{l} component of the propagation vector occurred between 0 and 2~T, accurately mirrored in the $+\tau$ and $-\tau$ reflections. Above 2~T the propagation vector remained approximately constant up to 10~T, hence indicating that the change in magnetic propagation is closely coupled to the magneto-electric polarization flip, particularly along the \textit{c}-axis direction. As both dipole and quadrupole dependences followed the same trend, one would predict a significant coupling between the manganese and terbium sub-lattices; this is contrary to that indicated by neutron diffraction measurements \cite{chapon04}. However, synchrotron x-ray diffraction provides far greater reciprocal space resolution and ionic and electronic band sensitivity, and so may reveal such subtle effects. This result therefore provides new evidence of the terbium ions coupling to the manganese magnetic structure under applied magnetic fields.

The integrated intensities of the two (0+$\delta$,~0,~5+$\tau$) resonances are given in the lower pane of figure \ref{fielddepfig}, clearly showing the suppression of the quadrupole resonance. The ratio between quadrupole and dipole intensities decreases monotonically until extinction at 2~T. There also occurs a decrease in the intensity of the dipole resonance; further evidence of a change in the manganese sub-lattice magnetic structure between 0 and 2~T, possibly due to  a coupling between the terbium $4f$ and manganese $3d$ states.

\section{Conclusions}

To summarize, upon applying an external magnetic field parallel to the \textit{a}-axis of \tmo\ in the LT-ICM phase, we observed a significant change in the relative intensities of the Tb $L_\mathrm{III}$, dipole and quadrupole resonant x-ray diffraction signals. The two directly probe the terbium $5d$ and $4f$ electronic states, respectively. The quadrupole resonance was shown to become suppressed at 2~T, coinciding with the electric polarization reversal. A degree of antiferromagnetic propagation of the $4f$ states remains above 2~T, common with the manganese magnetic structure. The terbium sub-lattice therefore adopts a canted antiferromagnetic structure with a large ferromagnetic component along the \textit{a}-axis. This scenario explains the reduced saturation magnetic moment of the terbium ions, evident in magnetization measurements. Furthermore, due to the extremely high reciprocal space resolution of synchrotron x-ray diffraction, we were able to show a coupling between the manganese and terbium sub-lattices, previously thought to be completely decoupled. Such a coupling is of key importance in understanding the possible magneto-electric mechanisms in these multiferroic materials.

\begin{acknowledgments}
RDJ, SRB and PDH would like to thank E.P.S.R.C. for funding. CHD is thankful to NSC of Taiwan for the research grant through the number 99-2112-M-032-005-MY3. We are grateful to the European Synchrotron Radiation Facility for the beam time and access to their facilities.
\end{acknowledgments}

\bibliography{tbmn2o5_axis_bib}

\end{document}